\documentclass[letterpaper,english,reprint,aps,prl]{revtex4-1}
\usepackage[T1]{fontenc}
\usepackage[latin9]{inputenc}
\usepackage{amsmath}
\usepackage{amssymb}
\usepackage{graphicx}
\usepackage{lineno}
\usepackage{hyperref}
\hypersetup{colorlinks,
citecolor=blue}
\usepackage{geometry}
\makeatletter

\pdfpageheight\paperheight
\pdfpagewidth\paperwidth

\makeatother

\usepackage{babel}
\begin{document}
\title{Exact solution for progressive gravity waves on the surface of
a deep fluid}
\author{Nail S. Ussembayev}
\email[]{nussemba@indiana.edu}

\affiliation{Computer, Electrical, and Mathematical Sciences and Engineering Divison~\\
 King Abdullah University of Science and Technology, Thuwal 23955-6900,
KSA }
\begin{abstract}
Gerstner or trochoidal wave is the only known exact solution of the
Euler equations for periodic surface gravity waves on deep water.
In this Letter we utilize Zakharov's variational formulation
of weakly nonlinear surface waves and, without truncating the Hamiltonian
in its slope expansion, derive the equations of motion for unidirectional
gravity waves propagating in a two-dimensional flow. We obtain an
exact solution of the evolution equations in terms of the Lambert
$W$-function. The associated flow field is irrotational.
The maximum wave height occurs for a wave steepness of 0.2034 which
compares to 0.3183 for the trochoidal wave and 0.1412 for the Stokes
wave. Like in the case of Gerstner's solution, the limiting wave of a new type has a cusp of zero angle at its crest. 
\end{abstract}

\maketitle

\title{An exact solution for progressive gravity waves on the surface of
a deep fluid}

\textit{Background.---}The water waves problem even in the completely
idealized setting of a perfect fluid is notoriously difficult from
a mathematical point of view. This is one of the reasons we have very
few explicit solutions to the fully nonlinear free-surface hydrodynamics
equations despite extensive research over hundreds of years. The first
exact solution was provided by Gerstner \cite{Gerstner}
in 1802 for pure gravity waves of finite height propagating on the
surface of an inviscid and incompressible fluid of infinite depth.
The practical importance of trochoidal waves became apparent when
in the 1860s Rankine and Froude independently rediscovered them in
connection with the ship-rolling problem \cite{Rankine}. The fact that these waves
are rotational and cannot be generated from the fluid at rest was
seen as somewhat unsatisfactory. Moreover, since the fluid particles
move on a closed circular orbit, Gerstner's solution does not predict the
observed mass transport in the direction of wave propagation \cite{Wiegel}.

In 1847 Stokes presented an approximate theory for irrotational waves
and showed that for deep water the wave profile coincides with a trochoid
to the third order, but not to the fourth \cite{Stokes}.
He also conjectured the existence of the wave of greatest height distinguished
by a sharp edge and argued that the enclosing angle must be $120^{\circ}$.
The mere existence of such waves was rigorously verified only forty
years ago \cite{Toland}, clearly illustrating the
inherently complicated structure of the underlying nonlinear equations.
The existence of extreme waves that are, as hypothesized by Stokes,
strictly convex between successive crests was established very recently,
but in spite of some numerical evidence supporting the convexity of
all waves of extreme form their uniqueness remains an open problem
\cite{Plotnikov}.

Apart from Crapper's exact solution for pure capillary waves  \cite{Crapper} found
in 1957  and some explicit Gerstner-type solutions for equatorially
trapped waves discovered lately \cite{Constantin},
the field of exact solutions for water wave equations has not experienced
significant advancement since the last century. In this Letter, we
present an exact solution for planar periodic waves of finite amplitude
moving uniformly and without change of form on a fluid of unlimited
depth when only gravity acts as a restoring force. Following Zakharov \cite{Zakharov},
we use the power of Hamiltonian formalism to write an asymptotic expansion
of the equations of motion for a pair of canonically conjugate variables
-- the surface elevation and the velocity potential evaluated
on the free surface. The expansion coefficients simplify remarkably
for two-dimensional waves propagating in the same direction and the
infinite series can be summed in a closed form if the wave steepness
is small. From the stationary periodic solution of the resulting nonlinear
equations, an exact irrotational solution to the full system of hydrodynamic
equations with a free surface is then recovered. The solutions are
provided in Eulerian coordinates \cite{fn1}. 

Exact solutions play an important role not just because all the properties
of the motion can be conveniently expressed and analyzed without any
approximation, but also because they can provide a key insight into
the nature of more complex and physically more realistic flows, say,
in which waves propagate under the combined influence of gravity and
surface tension.

Let us finish this introduction with a note that two-dimensional flows
are not of purely theoretical interest. Their fundamental importance
stems from the observation that in many situations the motion of water
waves can be treated as quasi-two-dimensional. For example, waves
generated in the open ocean while coming ashore often appear as a
wave-train of very long nearly parallel crests \cite{Johnson}
and due to this symmetry can be more or less accurately studied and
simulated by various two-dimensional models (see, e.g., \cite{Chalikov}). 

\textit{Theoretical formulation.---}The potential flow $\phi(x,y,t)$
of an ideal incompressible fluid in the region $\Omega(t)=\left\{ (x,y)\in\mathbb{R}^{2}\colon-\infty<y\leq\xi(x,t)\right\} $
bounded by the free surface $\Gamma(t)=\left\{ y=\xi(x,t)\right\} $
satisfies the following set of equations: 

\begin{subequations}\label{eq1}
\begin{align}
\xi_{t}+\xi_{x}\phi_{x}-\phi_{y}=0, &  &  & \mbox{ on }\Gamma(t)\label{eq1a}\\
\phi_{t}+\frac{1}{2}\left(\phi_{x}^{2}+\phi_{y}^{2}\right)+g\xi=0, &  &  & \mbox{ on }\Gamma(t)\label{eq1b}\\
\phi_{xx}+\phi_{yy}=0, &  &  & \mbox{ in }\Omega(t)\label{eq1c}\\
\phi_{y}\to0, &  &  & \mbox{ as }y\to-\infty\label{eq1d}
\end{align}
\end{subequations}
where $g$ is the constant gravitational acceleration in the negative
$y$ direction and subscripts denote partial derivatives. The effects
of surface tension are neglected and the pressure at the free surface
is taken to be zero. Once $\xi(x,t)$ and $\phi(x,y,t)$ are found,
the velocity field and the pressure within the fluid domain follow
from 

\begin{align*}
 & {\bf v}(x,y,t)=\nabla\phi,\\
 & p(x,y,t)=-\rho\left(gy+\phi_{t}+\frac{|{\bf v}|^{2}}{2}\right)
\end{align*}
where $\rho$ is the fluid density. Since $\phi$ is harmonic in $\Omega(t)$,
it is uniquely determined by its trace on the free boundary, $\psi(x,t)=\left.\phi\right|_{y=\xi(x,t)}$,
and \eqref{eq1d}. Zakharov ingeniously realized that equations  \eqref{eq1a}-\eqref{eq1b}
describing the wave motion can be put in the form of Hamilton's
canonical equations 
\begin{equation}\label{EOM}
\rho\frac{\partial\xi}{\partial t}=\frac{\delta H}{\delta\psi},\ \ \rho\frac{\partial\psi}{\partial t}=-\frac{\delta H}{\delta\xi}
\end{equation}
with the Hamiltonian given as the sum of the potential energy measured
with respect to the undisturbed fluid level $y=0$ and the kinetic
energy \cite{Zakharov}: 
\begin{equation}\label{Ham}
H=\frac{\rho}{2}\int_{-\infty}^{\infty}dx\left(g\xi^{2}+\int_{-\infty}^{\xi}|{\bf v}|^{2}dy\right).
\end{equation}
The Laplace equation \eqref{eq1c} and the decay condition at infinity \eqref{eq1d} represent
the constraints on the dynamical system whose total energy $H$ is
a constant of motion. 

The major difficulty in applying the Hamiltonian formalism for water
waves lies in expressing the kinetic energy in terms of free surface
variables, $\xi$ and $\psi$, alone. When the wave steepness is small,
$|\xi_{x}|\ll1$, one can write the expansion of the Hamiltonian in
a power series in the amplitudes of the waves. Let $\hat{\xi}_{k}=\hat{\xi}_{-k}^{*}$
and $\hat{\psi}_{k}=\hat{\psi}_{-k}^{*}$ denote the Fourier components
of $\xi$ and $\psi$, then upon solving the boundary-value problem for Laplace's equation the Hamiltonian \eqref{Ham} reads

\begin{align}\label{HamF}
&H=\frac{1}{2}\int|k|\hat{\psi}_{0}^{*}\hat{\psi}_{0}+g\hat{\xi}_{0}^{*}\hat{\xi}_{0}dk +\nonumber \\ 
 &+\sum_{n=1}^{\infty}\int E_{0,1,2,\dots,n+1}^{(n+2)}\hat{\psi}_{0}\hat{\psi}_{1}\prod_{i=2}^{n+1}\hat{\xi}_{i}\delta\left(\sum_{i=0}^{n+1}k_{i}\right)\prod_{j=0}^{n+1}dk_{j}. 
\end{align} where we set $\rho=1$ \cite{fn2}. Computing the expansion kernels $E_{0,1,2,\dots,n+1}^{(n+2)}$ relies on a series reversion associated with expressing the velocity potential in terms of its value at the free surface \cite{Krasitskii}. These kernels can be obtained in principle to all orders, however,  the iteration process quickly becomes cumbersome. 
A fully explicit recursion relation for computing $E_{0,1,2,\dots,n+1}^{(n+2)}$ without a laborious series reversion has been proposed recently in \cite{Ussemb}
for an arbitrary depth and arbitrary space dimension. 

The Fourier transform is a canonical transformation, and hence the evolution equations
\eqref{EOM} preserve their Hamiltonian form in the canonically conjugate variables
$\hat{\xi}_{k}$ and $\hat{\psi}_{k}$: \begin{widetext}
\begin{subequations}\label{EOMF}
\begin{align}
 & \frac{\partial\hat{\xi}_{k}}{\partial t}=\frac{\delta H}{\delta\hat{\psi}_{k}^{*}}=|k|\hat{\psi}_{k}+2\sum_{n=1}^{\infty}\int E_{-0,1,2,\dots,n+1}^{(n+2)}\hat{\psi}_{1}\prod_{i=2}^{n+1}\hat{\xi}_{i}\delta\left(k-\sum_{i=1}^{n+1}k_{i}\right)dk_{12\dots n+1},\label{eq5a}\\
 & \frac{\partial\hat{\psi}_{k}}{\partial t}=-\frac{\delta H}{\delta\hat{\xi}_{k}^{*}}=-g\hat{\xi}_{k}-\sum_{n=1}^{\infty}\int nE_{1,2,\dots,n+1,-0}^{(n+2)}\hat{\psi}_{1}\hat{\psi}_{2}\prod_{i=3}^{n+1}\hat{\xi}_{i}\delta\left(k-\sum_{i=1}^{n+1}k_{i}\right)dk_{12\dots n+1}\label{eq5b}
\end{align}
\end{subequations}
\end{widetext}
where \eqref{HamF} was used to compute the variational derivative.
The expansion kernels posses a remarkable property revealed by the
recursion relation: for any fixed $n\in\mathbb{N}$, it holds that
\begin{equation}\label{ker1}
E_{-\sum_{i=1}^{n+1}k_{i},k_{1},k_{2},\dots,k_{n+1}}^{(n+2)}=0
\end{equation}
if all wavenumbers have the same sign  \cite{Ussemb} and
\begin{equation}\label{ker2} E_{k_{1},k_{2},\dots,k_{n+1},-\sum_{i=1}^{n+1}k_{i}}^{(n+2)}=-\frac{\sigma^{n+1}}{n(2\pi)^{n/2}}\prod_{i=1}^{n+1}k_{i}
\end{equation}
with $\sigma=-1$ ($\sigma=+1$) if all wavenumbers are positive
(resp. negative). Hence, by considering waves traveling in one direction
we can drastically simplify the equations of motion. Indeed, taking the inverse
Fourier transform of \eqref{EOMF} and using \eqref{ker1}-\eqref{ker2} we obtain 

\begin{subequations}\label{evol}
\begin{align}
 & \xi_{t}^{\pm}\pm i\psi_{x}^{\pm}=0\label{evol1}\\
 & \psi_{t}^{\pm}+g\xi^{\pm}=\frac{\left(i\psi_{x}^{\pm}\right)^{2}}{1\mp i\xi_{x}^{\pm}}\label{evol2}
\end{align}
\end{subequations}
where $f^{\pm}=P^{\pm}f$, $P^{\pm}=\frac{1}{2}\left(I\pm i\mathbb{H}\right)$
are orthogonal self-adjoint projections onto positive and negative
wavenumber components, $\mathbb{H}$ is the Hilbert transform \cite{fn3}.
In summary, the small-amplitude assumption, $|\xi_{x}|\ll1$, in conjunction with the unidirectionality of wave propagation for
which relations \eqref{ker1}-\eqref{ker2} are valid in one dimension allows us to write apparently intangible equations \eqref{EOMF} containing infinite sums in a concise form given in \eqref{evol}. 

A few remarks are in order. First, making a crude approximation $\left(1\mp i\xi_{x}^{\pm}\right)^{-1}=1+{\mathcal O}\left(\xi_{x}^{\pm}\right)$
in \eqref{evol2} and letting $g=0$ reduces the system \eqref{evol} to the equations (17)-(18) derived in
\cite{Kuznetsov} to study the formation of singularities on the free surface
of an ideal fluid in the absence of gravitational force. Second, differentiating \eqref{evol1} with respect to time and \eqref{evol2} with respect to space yields the conservation form
$$\left(\xi_{t}^{\pm}\right)_{t}\pm\left(\frac{i\left(\xi_{t}^{\pm}\right)^{2}}{1\mp i\xi_{x}^{\pm}}-ig\xi^{\pm}\right)_{x}=0$$ 
which uncouples the free surface elevation from $\psi^{\pm}$. Last, using \eqref{evol1} it is possible to rewrite \eqref{evol2} as 
$$\psi_{t}+g\xi+\frac{1}{2}\psi_{x}^{2}-\frac{1}{2}\frac{\left(\xi_{t}+\xi_{x}\psi_{x}\right)^{2}}{1+\xi_{x}^{2}}=0$$ 
and \eqref{evol1} itself as a non-local equation 
$$\int_{-\infty}^{\infty}dxe^{-ikx+|k|\xi}\left(i\xi_{t}-\mbox{sgn}(k)\psi_{x}\right)=0$$ valid for every value of $k\in\mathbb{R}$. Thus we can recover from \eqref{evol} the non-local formulation of Euler's equations due to Ablowitz, Fokas and Musslimani (c.f. \cite{AFM}, p. 326). The AFM formulation is particularly useful for
deriving various asymptotic approximations  including the nonlinear Schr\"{o}dinger equation describing the envelopes of waves in deep water.   

\begin{figure}
\includegraphics[width=7.5cm]{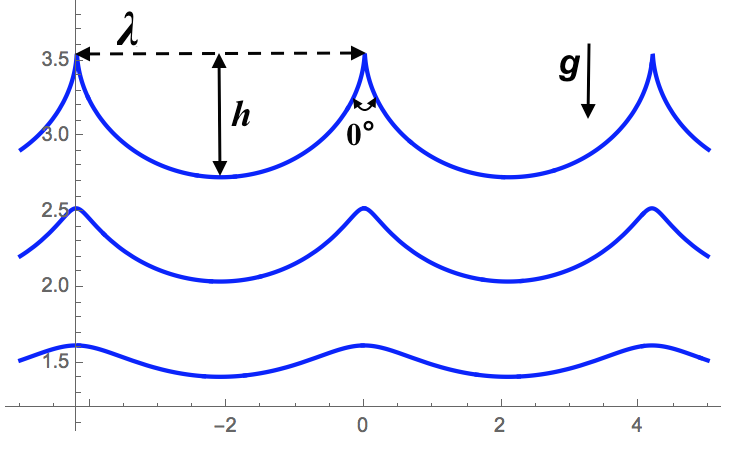}
\caption{Wave profiles (not to scale) for three values of the dimensionless
parameter $ka$ proportional to the wave slope: $ka=0.3678$ (top),
$ka=0.3$ (middle) and $ka=0.15$ (bottom). The crest loses its differentiability
when $ka=e^{-1}$. The initial phase is $\alpha=0$ and $g=1$.}\label{fig1}

\end{figure}

\textit{General solution.---}Next we attempt to find a periodic traveling
wave solution inserting the ansatz $f^{\pm}(x,t)=f^{\pm}\left(kx\mp\omega t\right)$
with $k>0$ to \eqref{evol}. The resulting system of ODEs is solved by the
Lambert $W$-function 
\begin{subequations}\label{sol}
\begin{align}
 & \xi^{\pm}(x,t)=-\frac{1}{k}W\left(-kae^{\pm i\left(kx\mp\omega t+\alpha\right)}\right)\label{sol1a}\\
 & \psi^{\pm}(x,t)=\frac{i\omega}{k^{2}}W\left(-kae^{\pm i\left(kx\mp\omega t+\alpha\right)}\right)+C\label{sol1b}
\end{align}
\end{subequations}
if $\omega$ verifies the deep water dispersion relation, i.e. $\omega=\sqrt{gk}$.
The two arbitrary constants in the solution of the equations of motion
are represented by an additive constant $C\in\mathbb{C}$ which we
can set to zero without loss of generality and the complex amplitude
$A=ae^{i\alpha}$ whose argument is the initial phase $\alpha$ and
the absolute value defines the first-order wave amplitude $a$ (see
\eqref{expan} below). The real part of $\xi^{+}$ ($\xi^{-}$ ) describes a
wave profile moving to the right (resp. left) with speed $c=\omega/k$
(see Fig. \ref{fig1}). One can check directly that $\phi(x,y,t)=-\frac{ia\omega}{k}e^{ky}e^{\pm i\left(kx\mp\omega t+\alpha\right)}$
satisfies the system \eqref{eq1}, is irrotational $\nabla\times{\bf v}=0$ \cite{fn4},
and equals $\psi^{\pm}(x,t)$ on the free surface since for any $z\in\mathbb{C}$
the Lambert function obeys the identity $z=W(z)e^{W(z)}$. Unless
otherwise stated, $W(z)$ will denote the principle branch of the
function. It has a branch cut along the negative real axis, ending
at $-e^{-1}$ \cite{Corless}. 

\begin{figure}
\includegraphics[width=7.5cm]{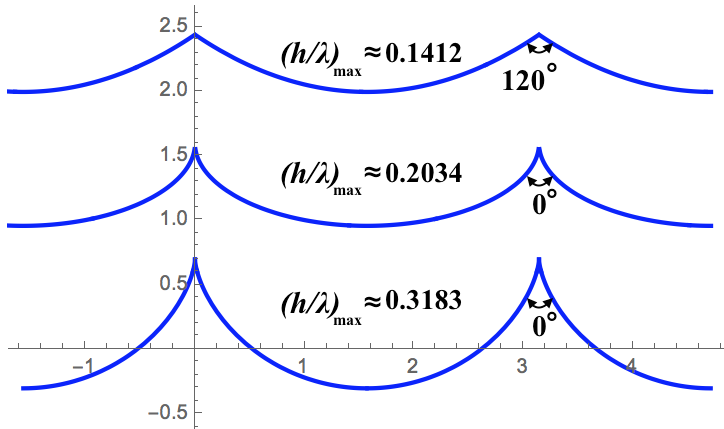}
\caption{Extreme wave profiles (not to scale) with the same horizontal distance
between two successive maxima: Stokes wave (top), real part of $\xi^{\pm}(x,t)$
(middle) and Gerstner wave (bottom). Stokes wave is plotted using
a close one-term piecewise approximation suggested in  \cite{Rainey}.
Limiting Gerstner wave is an upside-down cycloid.}\label{fig2}
\end{figure}

The wave profile changes from sinusoidal when $ka$ is small to the
wave with a sharp crest of angle $0^{\circ}$ when $ka=e^{-1}\approx0.3678$.
The position of the crest relative to the undisturbed water level
is $-k^{-1}W(-ka)$ and that of trough is $-k^{-1}W(ka)$ so that
the height of the wave is given by
\[
h=\max_{x\in\mathbb{R}}\xi^{\pm}-\min_{x\in\mathbb{R}}\xi^{\pm}=\frac{W(ka)-W(-ka)}{k}.
\]
The maximum wave height occurs for the wave steepness 
\[
\left(\frac{h}{\lambda}\right)_{\max}=\frac{1+W(e^{-1})}{2\pi}\approx0.2034
\]
where $\lambda=2\pi/k$ is the wavelength. This value is smaller than
the maximum wave steepness for the Gerstner wave, $\left(h/\lambda\right)_{\max}=\pi^{-1}\approx0.3183$
\cite{Wehausen}, but larger than that for the highest
Stokes wave in deep water, $\left(h/\lambda\right)_{\max}\approx0.1412$
\cite{Schwartz} as shown in Fig. \ref{fig2}. The free surface velocity potential
$\psi^{\pm}(x,t)$ is zero at the crest and trough of the wave and
develops a jump discontinuity at the crest as $ka$ approaches $e^{-1}$.
The horizontal velocity is highest at the crest of the wave and lowest
at the trough whereas the vertical velocity is zero at both these
locations. 

\textit{Discussion.---}Asserting the flow of physical interest to be
free of rotation, Lamb \cite{Lamb} and Stokes did
not endorse Gerstner's exact solution as much as naval architects
and engineers. In 1847 Stokes proposed an approximate solution for
irrotational motion by means of a perturbation series: he sought the
free-surface elevation as an infinite Fourier series $\sum_{n=1}^{\infty}a_{n}\cos\left(n(kx-\omega t)\right)$
with unknown coefficients $a_{n}$ and without discussing the convergence
properties. The convergence of Stokes' series for small-amplitude
deep water waves was proved by Levi-Civita \cite{Levi} in 1925.
A representation similar to the Stokes' expansion can be written for
the waves studied in this Letter. Indeed, for small values of the
parameter $ka$ using the Taylor series of the Lambert function we
obtain 
\begin{align}\label{expan}
\mbox{Re}\left(\xi^{\pm}(x,t)\right) & =a\cos\beta+ka^{2}\cos(2\beta)+\frac{3}{2}k^{2}a^{3}\cos(3\beta)\nonumber\\
 & +\frac{8}{3}k^{3}a^{4}\cos(4\beta)+O\left((ka)^{5}\right)
\end{align}
where $\beta=\pm\left(kx\mp\omega t+\alpha\right)$ with $\omega=\sqrt{gk}$.
The radius of convergence of the series for $W(z)$ around $z=0$
is precisely equal to $1/e$, i.e. to the value of $ka$ for the wave
of greatest height. One can compare expression \eqref{expan} to the corresponding
fourth-order Stokes expansion in deep water 
\begin{align*}
 \xi(x,t) &= a\cos\beta-\left(\frac{1}{2}ka^{2}+\frac{17}{24}k^{3}a^{4}\right)\cos(2\beta)+\\
 & +\frac{3}{8}k^{2}a^{3}\cos(3\beta)-\frac{1}{3}k^{3}a^{4}\cos(4\beta)+O\left((ka)^{5}\right)
\end{align*}
where the dispersion relation
involves the amplitude $\omega^{2}=gk(1+k^{2}a^{2}+\dots)$ (see, e.g., Lamb p. 419). 

Stokes theory is limited to small amplitude waves and due to the convergence
issues cannot yield the extreme wave for any value of the water depth.
Nevertheless, Stokes gave an elegant argument to show that if a cusp
is attained in an irrotational flow, then its tangents at the apex
necessarily make an angle of $120^{\circ}$. How does this agree with
our finding that the sharp-edged extreme wave has an included angle
of $0^{\circ}$ at the crest? To answer this question let us switch to a frame moving with
the crest and introduce polar coordinates $r$, $\theta$ with the origin
at the vertex, i.e. let $x+i(y+k^{-1}W(-ka))=re^{i\theta}$ where
$\theta$ is measured from one of the branches of the wave. In the
vicinity of the vertex the velocity potential behaves as $\phi\propto r^{n}\cos(n\theta)$
so that the tangential and normal velocity components are $v^{r}\propto r^{n-1}\cos(n\theta)$
and $v^{\theta}\propto r^{n-1}\sin(n\theta)$, respectively. In steady
flow, according to the Bernoulli equation we have $v^{2}\propto r$
which implies that $n=3/2$. The normal velocity must vanish for a
point on the wave profile and hence $3/2\theta=\pi$, giving the Stokes'
result. However, $\theta=120^{\circ}$ is not the only solution to
$v^{\theta}=0$ as the trivial solution $\theta=0^{\circ}$ also satisfies it.

\medskip The author is supported by the KAUST Fellowship.


\begin{thebibliography}{14}
\bibitem{fn1} Recall that Gerstner waves do not
have any explicit representation in Eulerian variables.
\bibitem{fn2} We use a compact notation in which $\hat\xi_j=\hat\xi(k_j,t)$, $\prod_{m=n}^l$ for $l<n$ is 1, $dk_{123}=dk_1dk_2dk_3$ etc., and subscript 0 corresponds to $k$. The star denotes complex conjugation, $\hat f(k)=\frac{1}{\sqrt{2\pi}}\int_{\mathbb{R}}{f}(x)e^{-i k x}dx$ is the Fourier transform of $f(x)$ and the Dirac delta function is $\delta(k)=\frac{1}{2\pi}\int_{\mathbb{R}} e^{ikx}dx$.
\bibitem{fn3} The Hilbert transform is defined by $\left(\mathbb{H}f\right)(x)=\frac{1}{\pi}\mbox{p.v.}\int_{\mathbb{R}}\frac{f(x')}{x-x'}dx'$, $\mathbb{H}e^{ikx}=-i\mbox{sgn}(k)e^{ikx}$.
\bibitem{fn4} In a two-dimensional setting the vorticity vector is identified with its middle component which is orthogonal to the plane of the flow.
\end{thebibliography}
\end{document}